\documentclass[preprint]{aastex}
\usepackage{emulateapj5}
\def\stacksymbols #1#2#3#4{\def\theguybelow{#2}
        \def\verticalposition{\lower#3pt}
        \def\spacingwithinsymbol{\baselineskip0pt\lineskip#4pt}
        \mathrel{\mathpalette\intermediary#1}}
\def\intermediary #1#2{\verticalposition\vbox{\spacingwithinsymbol
        \everycr={}\tabskip0pt
        \halign{$\mathsurround0pt#1\hfil##\hfil$\crcr#2\crcr
                \theguybelow\crcr}}}
\def\lta{\stacksymbols{<}{\sim}{2.5}{.2}}
\def\gta{\stacksymbols{>}{\sim}{3}{.5}}

\begin{document}

\title{RAPID COOLING OF DUSTY GAS IN ELLIPTICAL GALAXIES}

\author{William G. Mathews\footnotemark[1] 
\& Fabrizio Brighenti{\footnotemark[1]$^,$\footnotemark[2]}}

\footnotetext[1]{UCO/Lick Observatory,
Dept. of Astronomy and Astrophysics,
University of California, Santa Cruz, CA 95064 mathews@ucolick.org}

\footnotetext[2]{Dipartimento di Astronomia,
Universit\`a di Bologna,
via Ranzani 1,
Bologna 40127, Italy brighenti@bo.astro.it}






\vskip .2in

\begin{abstract}

We propose a stellar origin for the central dust clouds 
observed in most giant elliptical galaxies.
Dusty gas ejected from evolving red giant 
stars in E or cD galaxies
can cool rapidly even after 
entering the hot, X-ray emitting gas.
Cooling by thermal collisions with dust grains 
can be faster than either the dynamical time in the galactic potential 
or the grain sputtering time. 
Some grains survive in the cooled gas.
Dusty stellar outflows cool more efficiently 
in the central regions 
where the stellar metallicity is higher.
Mergers with gas and dust-rich dwarf galaxies 
may occasionally occur but are not required to 
explain the observed dust clouds.
\end{abstract}


\section{Introduction}

While most of the interstellar gas mass in elliptical galaxies, 
typically $\sim 10^9$ $M_{\odot}$, is at 
the galactic virial temperature $T_{vir} \sim 10^7$ K, 
the cores of $\sim 80$ percent of all large E galaxies
contain dusty clouds of cold gas 
of mass $\lta 10^4 - 10^5$ $M_{\odot}$ 
(e.g. van Dokkum \& Franx 1995). 
In addition, diffuse optical line
emission from gas at $T \sim 10^4$ K 
is observed in most or all elliptical galaxies
within a few hundred parsecs from the center 
(e.g. Caon et al. 2000), 
particularly in those galaxies formerly thought to contain 
``cooling flows.'' 
The notion that the hot gas cools to low temperatures 
is not supported by XMM spectra
(e.g. Peterson et al. 2001; Xu et al. 2001), 
i.e., the observed cooling rate
${\dot M}$ is lower than that expected in traditional 
cooling flows, $\sim 1$ $M_{\odot}$ yr$^{-1}$.
In any case, cooling of essentially dust-free hot gas 
cannot produce dust clouds.
In the following we propose that the dusty clouds 
are formed from gas expelled from evolving 
metal-rich red giant stars in E galaxy cores. 
Even if these dusty stellar envelopes 
are heated to $\sim T_{vir}$ after leaving the stars,
the gas can cool rapidly by collisions 
between thermal electrons and dust grains.

\section{Cooling-Sputtering Evolution}

Direct evidence that mass is being lost from evolving stars in 
X-ray luminous ellipticals is provided by 
near-IR ($\sim 10\mu$m) observations of hot circumstellar dust
(Knapp et al. 1989, 1992; Athey et al. 2002).
The 9.7 $\mu$m peak from (oxygen-rich) 
silicate grains apparent in ISO 
observations places an upper limit on the grain radius, 
$a \le 1$ $\mu$m (Figure 5 of Laor \& Draine 1993).
According to Athey et al., the total luminosity of  
excess near-IR radiation is consistent with 
expected total stellar mass loss rates in giant E galaxies,
$\sim 1$ $M_{\odot}$ yr$^{-1}$.

However, as dusty gas moves away from orbiting, mass-losing 
red giant stars, 
it encounters the hot interstellar gas at 
$T \sim T_{vir} \sim 1$ keV 
and is violently disrupted by hydrodynamic instabilities.
The interaction of the stellar wind/planetary nebula with 
the hot gas is extremely complicated, 
but this ejected gas cannot remain at $T \sim 10^4$ K
for more than $t_{H\alpha} \sim 10^5$ years without exceeding the 
typical H$\alpha$ + [NII] luminosities observed
(Mathews \& Brighenti 1999).
XMM spectra show no evidence for slowly heating gas 
at $T \lta T_{vir}/2$.
This empirical evidence supports rapid thermal mixing 
of the stellar ejecta, which is probably a consequence of 
gasdynamical instabilities that enormously increase 
the surface area between the stellar ejecta and the 
ambient hot gas. 
In view of the complexity of this interaction, the thermal 
evolution $T(t)$ of gas ejected from stars is difficult to 
estimate, but limits can be set.
In the following discussion we adopt a useful limiting 
assumption for $T(t)$: 
the dusty gas is rapidly heated to the temperature of 
the local hot gas with the dust intact. 
This 
assumption maximizes the likelihood that the stellar 
gas will remain thermally merged with the hot gas phase.
Grain sputtering during the rapid heating process is 
unlikely since the heating time must be $\lta t_{H\alpha}$,
which is very much less than the sputtering time (see below).

When dust grains 
are introduced into a hot plasma with $T \gta 3 \times 10^6$ K, 
they are sputtered (eroded) by impacting 
thermal H$^+$ and He$^{+2}$ ions. 
The grain radius $a$ decreases at a rate
\begin{equation}
{da \over dt} = - n_p  h_w [ 1 + 
(T_s / T)^{2.5} ]^{-1}
~~~{\rm cm}~{\rm s}^{-1}
\end{equation}
(Draine \& Salpeter 1979; Tsai \& Mathews 1995)
where $n(He^{++}) = n(H^+)/10$.
Here $h_w = 3.2 \times 10^{-18}$ cm$^4$ s$^{-1}$ and 
$T_s = 2 \times 10^6$ K are fitting parameters, 
$n_p = n_e(4-3\mu)/(2+\mu)$ is the proton density and 
$\mu = 0.61$ is the molecular weight.
We assume for simplicity that the cooling occurs 
at constant gas pressure $P$ so 
$n_e = (2+\mu) P /5 \mu kT$.

As soon as the dusty gas is heated 
(by assumption) to the local 
gas temperature, it will begin to cool.
Cooling occurs by 
thermal X-ray emission from the gas and 
by inelastic impacts of thermal electrons with the grains.
Grain-electron 
cooling can greatly exceed that from thermal X-ray emission 
(e.g. Dwek 1987; Dwek \& Arendt 1992) 
provided the grains are not sputtered away.
If the cooling time is shorter than the dynamical time in the 
galactic potential, the gas will tend to cool 
approximately {\it in situ}, the case we consider here.
During isobaric cooling, the gas temperature decreases 
according to 
\begin{equation}
{dT \over dt} = - {2 \over 5} {T \over P} 
{\rho^2 \over m_p^2} (\Lambda + \Lambda_d)
= - {2 \over 5} \left(\mu \over k\right)^2
P ~ {\Lambda + \Lambda_d \over T}.
\end{equation}
Here $\Lambda(T,z)$ is the usual thermal radiation 
coefficient for gas with 
metallicity $z$ in solar units (Sutherland \& Dopita 1993)
and $\Lambda_d$ is the coefficient for dust-induced cooling.

The cooling of the hot plasma by grains is due mostly to 
inelastic impacts of thermal electrons with the grains. 
For high temperature plasmas, such as we consider, 
the rate of electron-grain
collisions is unaffected by the grain charge 
(Draine \& Salpeter 1979) which we assume is regulated 
by emission of low energy electrons 
from the grain surface. 
An individual grain cools the local gas at a rate
\begin{equation}
{\dot e}_g = 4 \pi a^2 (1/4) n_e {\bar v}_e {\bar E}_e
= 1.27 \times 10^{-7} n_e T_7^{3/2} a_{\mu}^2
                        ~~~{\rm erg}~{\rm s}^{-1}
\end{equation}
where ${\bar E}_e = (3/2) k T$ and
${\bar v}_e = (8 kT / \pi m_e)^{1/2}$ are the
mean energy and velocity
of electrons, $T_7 = T/(10^7~{\rm K})$ and 
$a_{\mu}$ is the grain radius in microns.
At pressures of interest, 
$P_{-10} = P/(10^{-10}~{\rm dy})$, 
the plasma mean free path, $\lambda \approx 0.02 T^3/P_{-10}$ cm, 
is small; only electrons in the cooling regions 
collide with the grains; 
electrons from the hot ambient gas thermalize in the 
cooling plasma before colliding with a grain.
In NGC 4472, a giant elliptical in the Virgo cluster 
which we use for reference, 
$n_e \approx 0.0551 r_{kpc}^{-1.18}$ cm$^{-3}$ so
${\dot e}_g = 7.7 \times 10^{-9}a_{\mu}^2$ and 
$7.3 \times 10^{-10}a_{\mu}^2$ at $r = 1$ kpc 
and the effective radius $r = R_e = 8.57$ kpc respectively.
We adopt a distance of 17 Mpc to NGC 4472.

Grains can internally absorb the energy
of thermal electrons provided 
${\bar E}_e \lta E_* = 23 a_{\mu}^{2/3}$ keV
(Dwek \& Werner 1981), which we assume is generally
satisfied.
Cooling by electron impact on grains during 
the sputtering time, 
$t_{sp} \approx a / |da/dt| \approx 1.2 \times 10^8 a_{\mu} 
(n_e/10^{-2})^{-1}$ yrs, 
greatly exceeds the additional gas cooling
required to fully ionize the grain atoms
ejected in the sputtering process or to bring these 
sputtered ions up to local thermal velocities.

At any moment the space density of interstellar 
grains $n_g$ in luminous elliptical galaxies 
is very inhomogeneous, being concentrated 
near mass-losing stars. 
If $m_{ej}$ is the typical total mass of 
the stellar envelope ejected by evolving stars 
and $\rho_g = 3.3$ gm cm$^{-3}$ is 
the density of (silicate) grains, 
then the total number of grains expelled by a single star is 
${\cal N}_g = \delta m_{ej} / \rho_g (4/3) \pi a^3$
where $\delta$ is the ratio of the dust mass to the 
gas mass when ejected from the stars. 
We assume that $\delta = 0.01 z_{*,Fe}$ 
decreases with stellar metallicity 
as in NGC 4472 where
$z_{*,Fe} \approx 0.675 (r/R_e)^{-0.207}$
in solar units (Brighenti \& Mathews 1999).

The initial space density of grains is proportional 
to the local gas density   
$n_{g0} = ({\cal N}_g / m_{ej})\rho$. 
As the gas cools isobarically from some initial 
temperature $T_0$, 
the grain space density increases,  
$n_g = n_{g0} T_0/T 
= [\delta m_p/\rho_g (4/3) \pi a^3] (\mu P/k)/T$.
The dust cooling coefficient in Equation 2, 
$\Lambda_d = n_g {\dot e}_g /(\rho/m_p)^2$, is then
\begin{equation}
\Lambda_d 
=  {9 \over 8} { 2 + \mu \over 5 \mu}
\left( {8 \over \pi m_e} \right)^{1/2} 
{\delta m_p k^{3/2} \over \rho_g 10^{-4}}
~ {T^{3/2} \over a_{\mu}}~~~{\rm erg}~{\rm cm}^3~{\rm sec}^{-1}.
\end{equation}

To explore the influence of grain cooling we integrate 
Equations 1 and 2 beginning with gas at 
temperature $T_0 = T(r)$ and pressure $P(r)$ at 
galactic radius $r$ in the hot gaseous atmosphere 
of NGC 4472. 
The pressure $P = n_e(r) kT(r) 5/(2+\mu)$ 
is constant during cooling.
For simplicity,
we assume the gas is filled with grains all 
having the same initial radius $a_0$. 
As the gas cools the sputtered debris from the grains 
increases the metallicity of the gas phase.
To evaluate the plasma cooling rate coefficient $\Lambda(T,z)$ 
during the cooling, we assume that the gas phase 
metal abundance in solar units 
increases with decreasing grain mass ($\propto a^3$),
$z/z_0 = 1 - (a / a_0)^3 (\delta /\delta_{max})$, 
where $\delta_{max} \approx 0.053$ is the maximum ratio 
of dust to gas mass if all elements heaver than 
He were in grains.

\section{Results}

We describe solutions of Equations 1 and 2 
for the elliptical galaxy NGC 4472 with initial conditions at 
radii $r = 1$ and $r = R_e = 8.57$ kpc. 
According to our assumption, the dusty gas expelled from 
a star is rather quickly heated to 
the local temperature of the hot gas at these two 
galactic radii, $T_0 = 0.95 \times 10^7$ and 
$1.16 \times 10^7$ K, where the gas density is 
$n_{e0} = 6.52 \times 10^{-2}$ and $4.45 \times 10^{-3}$ 
cm$^{-3}$ respectively.
Figures 1a and 1b show the combined evolution of the gas 
temperature $T(t)$ and grain radius $a(t)$ for grains of initial 
radius $a_0 = 1$ and 0.1 micron, assuming $\delta = 0.01$
and $\delta = 0.0064$ at $r = 1$ and $r = 8.57$ kpc 
respectively.
For comparison we also plot the cooling curve $T(t)$ of 
dust-free gas ($a_0 = 0\mu$) with the same $T_0$ and $n_{e0}$. 

Dust-enhanced cooling can reduce the 
cooling time of recently ejected gas 
by over an order of magnitude compared to the dust-free case.
For fixed $\delta$, the ratio $\Lambda_d/\Lambda$ increases 
rapidly with decreasing grain radius 
$a_0$ because of the larger total 
surface area of smaller grains.

A defining characteristic of cooling flows is that the
isobaric radiative cooling time 
$t_{cool,rad}  =  5 m_p k T/ 2 \mu \rho \Lambda 
\approx 10^8 r_{kpc}^{1.2}$ yrs (in NGC 4472) exceeds 
the dynamical time $t_{dy} \approx [r^3/G M(r)]^{1/2}$ 
in the galactic potential. 
Dust-free, moderately positive density perturbations
do not cool appreciably faster than unperturbed regions.
If overdense regions remain coherent, they 
oscillate radially in the hot gas atmosphere,
becoming underdense relative to the ambient gas 
during half their cycle (e.g. Loewenstein 1989).

The dust-rich regions we describe here are not initially 
in an overdense state, but become overdense 
as the cooling proceeds.
The dust-rich gas cools locally if the 
cooling time is less than the dynamical 
time at the two selected galactic radii,
$t_{dy} = 2.1 \times 10^6$ at $r = 1$ kpc 
and $1.2 \times 10^7$ yrs at 8.57 kpc. 
Figures 1a and 1b show that $t_{cool} < t_{dy}$ 
is possible at $r = 1$ kpc, 
so gas at small galactic radii can cool to low temperature
and also preserve a small amount of its original dust 
since $t_{cool} < t_{sp}$.
The cooling curves in Figures 1a and 1b
are independent of $m_{ej}$ and are 
therefore unchanged if the dusty envelope 
is disrupted or fragmented during 
its interaction with the hot gas. 
We show one case in Figure 1b in which dusty gas 
with $a_0 = 0.1$ $\mu$m begins cooling at $T_0/3$,
before fully thermalizing with the hot gas. 
The cooling time is significantly reduced, 
confirming our assumption that $T(t=0) = T_0$ maximizes 
the cooling time.

As dust-rich stellar winds flow away from 
orbiting red giant stars, the winds are  
decelerated by interactions with the local hot gas, 
but the grains, impelled by their 
higher momentum, can move into the ambient gas.
It is possible in principle that the ejected 
dust cloud occupies and 
cools a gas mass larger than that ejected from the star, $m_{ej}$.
If the cooled mass exceeds $m_{ej}$ by a factor 
$\omega$, then the corresponding cooling evolution is 
found simply by replacing $\delta$ in Equation 4 
with $\delta/\omega$.
In Figure 1c we show cooling curves for $\omega = 2$ 
at the same two galactic radii for grains 
of initial radius $a_0 = 0.1$ micron. 
Both cooling times exceed the dynamical time 
at the galactic radii considered,
so the cooling gas may have a more complex dynamical 
evolution that may inhibit the cooling.

\section{Discussion and Conclusions}

Stellar mass loss has been regarded as an important 
internal source of hot gas within group or cluster-centered 
E galaxies.
A $\sim 13$ Gyr old stellar population of total mass $M_{*t}$
typically expels 
${\dot M}_* \sim 1.5  (M_{*t}/10^{12}~M_{\odot})$ 
$M_{\odot}$ yr$^{-1}$.
In cluster-centered galaxies, such as M87 in Virgo
and NGC 4874 in the Coma cluster,
the gas temperature rapidly decreases in the
central $\sim 15$ kpc from the
virial temperature of the cluster (3 - 8 keV) to
the stellar virial temperature $\sim 1$ keV at $r \lta 3$ kpc 
(Molendi 2002; Vikhliin et al. 2001).
Detailed gasdynamical models of traditional cooling inflows 
in these cluster-centered galaxies indicate that this steep
temperature gradient can only be understood if
gas is being cooled by thermalization of stellar ejecta
(Brighenti \& Mathews 2002a).
The increase in the hot gas oxygen abundance toward 
the center of M87 
(Gastaldello \& Molendi 2002) 
is expected if mass lost from the SNII-enriched stars 
has thermally mixed into the hot ISM and if the stars are 
more O-rich than the Virgo cluster gas.
Nevertheless, we propose here that the dust component
can cool the stellar ejecta soon after 
it thermally merges
with the ambient hot gas, particularly at 
$r \lta 1$ kpc. 

We now review how dust-enhanced 
cooling applies to the issues discussed in the Introduction.

(1) {\it What is the source of interstellar dust
observed in the cores of $\sim 80$ percent of all large E galaxies?}
van Dokkum \& Franx (1995) and others describe
small central dust disks, lanes or clouds typically
a few 100 pc in size.
The small masses of dust $\lta 10^4 - 10^5$ $M_{\odot}$
in these cores can easily be produced in $\sim 10^8 - 10^9$ yrs
as dust-rich gas cools near the
galactic center with incomplete sputtering.
It is difficult to accurately estimate the 
rate that dust accumulates near the center without knowledge 
of the dust size distribution, the exact thermal history 
$T(t)$ of the stellar ejecta, or the star formation rate 
in the cold, dusty clouds.
Since Type Ia supernova remnants occupy a 
very small fraction of the 
interstellar volume at any time (Mathews 1990),
heating by Type Ia supernovae is unlikely to interfere with the 
cooling we describe here.
Heating by active galactic nuclei 
has often been suggested to explain why 
the hot gas in E galaxies fails to cool to low temperatures 
(e.g., Rosner \& Tucker 1989; 
Binney \& Tabor 1995; Brighenti \& Mathews 2002b). 
If such heating occurs, 
it must be gentle enough not to 
destroy the observed central dust clouds or this dust must 
be rapidly regenerated.

(2) {\it What is the source of the diffuse optical line
emission (H$\alpha$ + [NII])
observed in most or all cooling flow galaxies?}
The velocities of this diffuse emission are unrelated to 
stellar velocities (Caon et al. 2000).
In some E galaxies the 
optical line emission spatially correlates with 
X-ray features (e.g. Trinchieri \& Goudfrooij 2002)
or traces the perimeters but not the centers 
of X-ray cavities 
(McNamara, O'Connell \& Sarazin 1996; Blanton et al. 2001).
While these observations suggest that 
some gas cools to $\sim 10^4$ K 
from the hot phase, there is no evidence for this
in {\it XMM} X-ray spectra. 
However, even if dusty gas ejected from metal-rich 
stars is heated to $\sim T_{vir}$, it can quickly cool back to 
$\sim 10^4$ K where the cooling may be temporarily arrested
by absorption of galactic UV starlight
(Binette et al. 1994).
Exactly what happens next is unclear, but the dust
can help cool the gas to much lower temperatures if the
clouds become optically thick to UV radiation.
It has often been suggested that dusty gas at $T \sim 10^4$ K 
derives from mergers with gas-rich dwarf galaxies
(e.g. Caon et al. 2000; Trinchieri \& Goudfrooij 2002);
this can be verified
if the $10^4$ K gas is counter-rotating.
Nevertheless, cold, dusty gas can arise naturally 
from stars in E galaxy cores.

(3) {\it Why is the observed cooling rate
${\dot M}$ from X-ray observations so low?}
Xu et al. (2001) found from RGS XMM observations 
of elliptical galaxy NGC 4636 that X-ray lines  
expected from gas cooling near $T \sim 2 \times 10^6$ K 
are unusually weak, indicating a total cooling rate of 
${\dot M} \lta 0.30$ $M_{\odot}$ yr$^{-1}$ 
in $r \lta 2.5$ kpc (for $d = 17$ Mpc). 
FUSE observations of OVI lines in NGC 4636 
(emitted at $T \sim 3 \times 10^5$ K) indicate 
${\dot M} \approx 0.17 \pm 0.02$ $M_{\odot}$ yr$^{-1}$
within 1.2 kpc (Bregman et al. 2001).
Both cooling rates are less than the $\sim 1 - 2$
$M_{\odot}$ yr$^{-1}$ predicted in traditional cooling
flow models (e.g. Bertin \& Toniazzo 1995).

Rapid dust-assisted cooling may be 
relevant to this discrepancy in two ways: 
(1) by reducing the total rate that gas enters the hot phase, 
and (2) by reducing the X-ray emission from cooling thermal gas at 
subvirial temperatures.
In models of traditional cooling flows, 
gas ejected from stars is usually assumed to enter the  
hot interstellar gas.
The stellar mass loss rate in a giant E galaxy, 
$\sim 1 M_{\odot}$ yr$^{-1}$, is an important source of gas 
since it is comparable to the expected 
cooling rate of the hot gas. 
In gas dynamical models 
the cooling rate ${\dot M}$ near the center of the flow 
varies inversely with the 
specific rate of stellar mass loss, $\alpha_* = {\dot M}_*/M_*$.
This is true even when there is an extended reservoir of 
circumgalactic hot 
gas due to cosmic accretion onto the surrounding galaxy group.
Therefore, the apparent cooling rate ${\dot M}$ 
observed with XMM should be reduced approximately in proportion 
to the fraction of stellar gas that fails to enter the hot gas 
because of dust-assisted cooling. 
However, if more than $\sim 90$ percent of the stellar
ejecta fails to enter the hot phase,
our gasdynamical models indicate that
the density of hot interstellar gas is lowered sufficiently
to initiate a strong galactic wind driven by Type Ia energy.
The transition to wind flows is rather sudden.
Since the very low $L_x$ characteristic
of strong winds are not observed,
this sets a limit on the efficiency of dust-assisted cooling
described here.

If dusty gas from stars is indeed heated to $\sim T_{vir}$, 
XMM observations require that it not emit thermal X-rays at 
intermediate temperatures either as it is heated or as it cools 
afterward.
We argue above that the heating phase is likely to be rapid, 
$\lta t_{H\alpha} \sim 10^5$ yrs.
When heated to $\sim T_{vir}$, 
the stellar gas may undergo rapid dust-assisted cooling, 
during which the X-ray emission from intermediate temperatures 
is greatly reduced. 
The radiation expected from an X-ray line of emissivity 
$\varepsilon_{line}(T,\rho)$ ergs s$^{-1}$ gm$^{-1}$, 
$E_{line} = \int \varepsilon_{line} 
dT / |dT/dt|$ ergs gm$^{-1}$,
varies inversely with the cooling rate $|dT/dt|$,
which can be $\gta 10$ times larger than normal plasma cooling. 
The X-ray line emission is suppressed by a similar factor.

In dust-assisted cooling, much of the thermal energy in
the hot gas is radiated by dust in the far-IR.
The maximum FIR luminosity expected,
$L_{FIR} \sim 5 {\dot M} k T/2 \mu m_p \sim
2 \times 10^{41} (T/10^7)({\dot M}/M_{\odot}~{\rm yr}^{-1})$
erg s$^{-1}$,
is consistent with ISO FIR luminosities observed in
NGC 4472 and NGC 4636, $\nu L_{\nu} \lta 2 - 5 \times 10^{41}$
erg s$^{-1}$ (Temi et al. 2003).

Nevertheless, it seems unlikely that dust-enhanced 
cooling can explain the weakness of 
X-ray cooling lines from gas at subvirial temperatures 
in clusters (e.g. Peterson et al. 2001) 
where the expected cooling rates, 
${\dot M} \gta 100$ $M_{\odot}$ yr$^{-1}$, greatly exceed 
the stellar mass loss rate in the central E or cD galaxy.

\vskip.4in
Studies of the evolution of hot gas in elliptical galaxies
at UC Santa Cruz are supported by
NASA grants NAG 5-8049 \& ATP02-0122-0079 and NSF grants
AST-9802994 \& AST-0098351 for which we are very grateful.
FB is supported in part by grants MURST-Cofin 00
and ASI-ARS99-74.


\vskip.1in
\figcaption[dustcoolfig1.ps]{
{\it (a):} Evolution of the gas temperature 
({\it solid lines}) and grain radius in $\mu$ 
({\it short dashed lines}) 
for dust-induced cooling in NGC 4472 at the effective 
radius, $r = 8.57$ kpc. 
Each set of curves is labeled with the initial grain radius
$a_0$ in $\mu$m 
and the solid curve for $a_0 = 0$ is the temperature evolution 
for dust-free gas at the same radius.
The ratio of the rates of dust and gas cooling $\Lambda_d/\Lambda$ 
is shown with {\it long dashed lines}.
The vertical heavy line 
shows the dynamical time $t_{ff}$ for freefall 
from $r$ to the galactic center.
{\it (b):} Same evolution as in {\it (a)} but at 
radius $r = 1$ kpc in NGC 4472. 
The heavy lines show the $T,a$-evolution of $a = 0.1$ $\mu$ grains 
if cooling commences at a lower temperature $T_0/3$.
{\it (c):} Two sets of evolutionary curves 
when the grains are dispersed into a mass of 
gas equal to twice that ejected from the star.
Both sets of curves are for $a_0 = 0.1$ $\mu$m and are 
labeled with the initial galactic radius.
\label{fig1}}

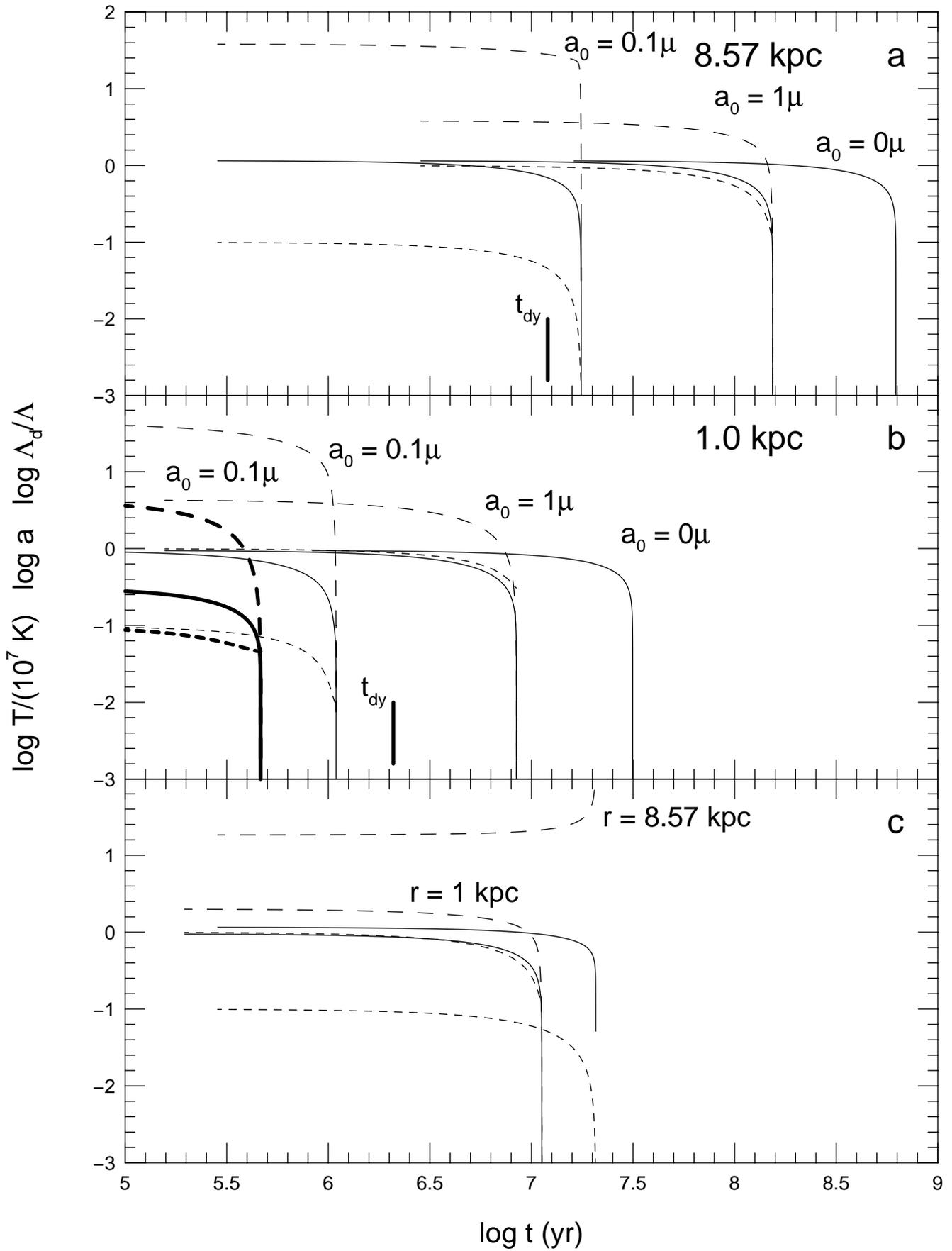
\end{document}